\documentclass[conference, letterpaper]{IEEEtran}
\usepackage{amssymb}
\usepackage{graphicx}
\usepackage{graphics}
\usepackage{subfig}
\newcommand{\ket}[1]{\left|{#1}\right\rangle}

\begin{document}

\title{Optimizing qubit Hamiltonian parameter estimation algorithms using PSO}

\author{\IEEEauthorblockN{Alexandr Sergeevich and Stephen D. Bartlett}
\IEEEauthorblockA{Centre for Engineered Quantum Systems, School of Physics,\\ University of Sydney, Sydney, NSW 2006, Australia}}

\maketitle

\begin{abstract}
We develop qubit Hamiltonian single parameter estimation techniques using a Bayesian approach. The algorithms considered are restricted to projective measurements in a fixed basis, and are derived under the assumption that the qubit measurement is much slower than the characteristic qubit evolution.  We optimize a non-adaptive algorithm using particle swarm optimization (PSO) and compare with a previously-developed locally-optimal scheme. 
\end{abstract}

\IEEEpeerreviewmaketitle

\section{Introduction}

The field of quantum computation is one of the fastest developing areas in modern physics. In order to perform quantum computation one needs to have a full and precise information about the parameters of quantum bit evolution in the set-up used. This makes quantum Hamiltonian parameter estimation one of the crucial problems in this area. 

The most common approach involves repeatedly initiating the system in some state and then performing a measurement after some time when the system is allowed to evolve under its Hamiltonian \cite{NC}. The choice of input states, evolution times, measurement bases and method of information treatment defines the estimation algorithm. Various approaches to the problem were suggested, including quantum process tomography \cite{Tom1, Tom2} and Fourier analysis \cite{cole}. Quantum process tomography requires a full control over state preparation and measurement which could be a hard task. In addition, often the preparation is done by using strong measurement in a certain basis which one can change only using evolution of the quantum bit (qubit) itself which is impossible without already knowing the Hamiltonian parameters. Thus finding algorithms which are able to work using a fixed basis is required. Fourier analysis method is inspired by classical signal processing. It involves repeated measurements after longer and longer evolution times and signal frequency reconstruction using inverse Fourier transform. However, it requires a large number of measurements to achieve moderate precision. I.e., in \cite{cole} it took $10^6$ measurements to achieve $10^{-3}$ joint variance in a two-parameter qubit Hamiltonian estimation. 

In order to completely use the information of each measurement it is necessary to use a Bayesian approach. It consists of representing the knowledge of an unknown parameter in terms of probability distribution and updating it according to the sequence of measurement outcomes. This approach is closely related to phase estimation techniques, which however also employ arbitrary states preparation \cite{NC, QPEA}.  

The particular problem considered in this paper is inspired by the developments in 2-electron double quantum dot (DQD) systems which can be successfully used for creating quantum bits and performing gate operations, in particular GaAs double quantum dots \cite{Reilly, Foletti}. In such systems qubit evolution is driven by a difference in $z$-component of magnetic field $\Delta B$ between the parts of the DQD. This field is due to the hyperfine interaction of the electrons' spins with atoms nearby lattice and so it is not stable and can't be controlled. Precise and quick measurement of this field is therefore crucial for quantum processing. One of the important limitations imposed by this system is a slow and complicated measurement process which requires considerably more time than the evolution ($\sim 6 \mu$s vs several nanoseconds). In order to optimize the estimation procedure we'll quantify the resources which are needed to be minimized in terms of number of measurements (not the total evolution time as in previous research available, i.e. \cite{WeismanNature}). 

The optimization process for such algorithm is hard as in general it requires sampling the probability distribution and browsing through an enormous amount of algorithm variants. The dependence of the final estimate variance (to be minimized) on the algorithm is a complicated function with many local minima and unpredictable behaviour. Particle swarm optimization (PSO) approach has proved to be useful in such situations when little is known about the function and global minimum is sought  \cite{KEPSO, PSObook}. It was also recently used in quantum phase algorithm optimization \cite{Hentschel1, Hentschel2}. In this paper we present a locally-optimal approach derived in \cite{Sergeevich} and demonstrate the improvement which can be achieved by using particle swarm optimization technique.

\section{Hamiltonian parameter estimation}

\subsection{Problem statement}
We consider the problem of single unknown parameter estimation of a qubit Hamiltonian of the form
\begin{equation}  
H=\frac{\omega\sigma_{z}}{2}, \label{Ham}
\end{equation}
where $\sigma_{z}$ is a $z$-Pauli matrix. 

The system is initiated in $\{\ket{-},\ket{+}\}$ basis (also referred to as "computational basis") by applying a projective measurement along $z$-axis. The time evolution of the state initiated as $\ket{-}$ under the Hamiltonian (\ref{Ham}) can be represented as
\begin{equation} 
\ket{\psi(t)} = \cos\left( \omega t/2 \right)\ket{-} - i\sin\left( \omega t /2\right)\ket{+}
\end{equation}
The measurement in the computational basis is then performed giving either $\ket{-}$ or $\ket{+}$ outcome.

We assume that the unknown parameter $\omega$ can be treated as a random variable uniformly distributed in $(0, \omega_0)$. The upper bound is defined by Nyquist-Shannon theorem as $\omega_0 = \pi/\Delta t$ where $\Delta t$ is the minimal accessible time interval. Using time intervals quantized as $t = m \Delta t, m \in \mathbb{N}$ it is easy to see that the evolution produces the following probabilities of the measurement outcomes:
\begin{equation}
p({+}|\omega) = \sin^2 \left(\frac{\pi m \omega}{2\omega_0} \right),  \;\;\;\;\; p({-}|\omega) = \cos^2
\left(\frac{\pi m \omega}{2\omega_0} \right). \label{prob}
\end{equation}

Assuming that each measurement takes the same amount of time, the problem would become to optimally select the sequence of evolution times $m_k \Delta t$.

\subsection{Bayesian approach}

Treating the information about $\omega$ as a probability distribution we now can use Bayesian updating to find the estimate. Taking the flat prior, distribution after some number of steps will merely be a normalized product of probability functions (\ref{prob}). The variance of this distribution can thus characterize the precision in parameter estimate. 

The choice of discrete evolution time selection as discussed above is essential for brute force algorithm optimization. In such case we can avoid sampling the distribution by representing it as a Fourier series. It can be done for any algorithm as $p({\pm}|\omega)$ are both harmonic functions. After $k$ measurements the distribution reads
\begin{equation}
  \label{sumo}
  P_{k} (\omega|r_k \ldots r_1) = \frac{1}{2} c_{k}(0)+\textstyle\sum\limits_{q=1} ^M c_{k}(q) \cos (q \pi\omega/\omega_0),
\end{equation}
where $M=\sum\limits_{j=1}^{k} m_j$ and $r_i$ are the measurement outcomes. The distribution is normalized by dividing
by $\frac{1}{2} \omega_0 c_{k}(0)$. The variance of this distribution $V = \langle\omega^2\rangle-\langle\omega\rangle^2$  can then be found as
\begin{equation}
  \label{variance}
  V = \frac{\omega_{0}^2}{3}+\sum_{q=1}^N\frac{2c_q\omega_{0}^2(-1)^q}{c_0(q  \pi)^2} 
  -\left [\frac{\omega_{0}}{2}+\sum_{q=1}^N\frac{c_q\omega_{0}[(-1)^q-1]}{c_0(q\pi)^2}\right ]^2.
\end{equation}

The expected variance $E[V]$ is now the function to minimize and it depends only on the selection of the set of evolution times. Also it is clear that $E[V]$ is constant under the permutations of elements of $\{m_i\}$ set.

\section{Algorithm optimization}

\subsection{Locally-optimal approach}

\begin{figure}[t]
\begin{picture}(300,167)
\put(20,3){\includegraphics[width=220pt]{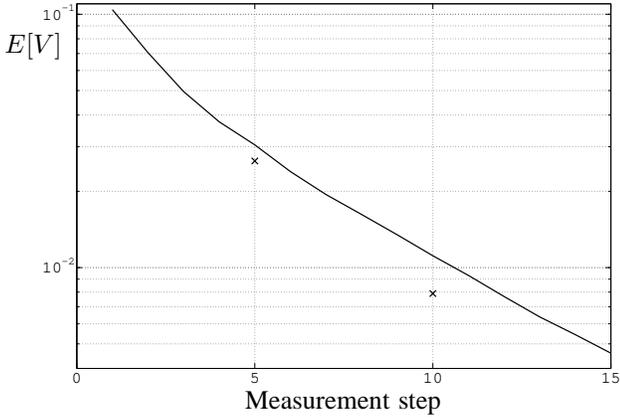}}
\put(6,137){$E[V]$}
\put(97,2){Measurement step}
\end{picture}
\caption{Performance of LONA algorithm (solid line) and improved algorithms found using PSO (crosses).}
\label{LONAfig}
\end{figure}

The straight way to optimize the algorithm is to use local optimization \cite{Sergeevich, WisMilBook}. This approach involves fixing first $k$ evolution times and optimizing the expected variance after the next measurement over the choice of $m_{k+1}$. Repeating this procedure for $k=1,...,N$ we create a locally-optimal non-adaptive (LONA) algorithm. First $10$ steps of LONA are shown in Table \ref{ALG10}. The performance of LONA algorithm is shown on Fig. \ref{LONAfig} as a solid line. 

\begin{table}[t]
\renewcommand{\arraystretch}{1.3}
\caption{LONA and optimized algorithm for 5 measurements.}\label{ALG5}
\centering
\begin{tabular}{c|c|c}
Step & LONA & PSO  \\ \hline
1 & 1 & 1.060 \\ \hline
2 & 1 & 1.082 \\ \hline
3 & 1 & 1.419 \\ \hline
4 & 2 & 2.138 \\ \hline
5 & 3 & 2.870 \\ 
\end{tabular}\end{table}

\begin{table}[t]
\renewcommand{\arraystretch}{1.3}
\caption{LONA and optimized algorithm for 10 measurements.}\label{ALG10}
\centering
\begin{tabular}{c|c|c}
Step & LONA & PSO  \\ \hline
1 & 1 & 1.071 \\ \hline
2 & 1 & 1.107 \\ \hline
3 & 1 & 1.161 \\ \hline
4 & 1 & 1.180 \\ \hline
5 & 1 & 1.200 \\ \hline
6 & 2 & 2.041 \\ \hline
7 & 2 & 2.152 \\ \hline
8 & 3 & 3.070 \\ \hline
9 & 4 & 3.970 \\ \hline
10 & 6 & 4.906 \\
\end{tabular}\end{table}

As it was argued in \cite{Sergeevich}, this approach demonstrates a significant improvement over a naive approach with fixed evolution time $m_k=1$ and Fourier techniques. However, numerical simulations show that this approach is not globally-optimal being slightly outperformed by $m_k = k$ algorithm in a short range around 60th measurement. In addition, as it was mentioned before, the general algorithm optimized over $m_k \in \mathbb{R}$ would be very complex computationally. Firstly, it will require distribution sampling, that is instead of a set of Fourier coefficients (which perfectly reconstruct the function), the distribution should be stored point-by-point with limited precision. Secondly, the brute force browsing through evolution times at each step and performing the averaging will make it unrealistic even for few steps as the number of operations grows as a very steep exponent. Globally-optimal optimization and selecting all the possible sets of $\{m_k\}$ that way becomes completely unrealistic.

\subsection{Particle swarm optimization}

The globally-optimal non-adaptive technique can be found with high precision employing PSO approach. It enables to perform the search in all the continuous space of the set of evolution times $\{t_i\}$. We are sampling the $\omega$ probability distribution using $10^4$ points which gives $\lesssim 10^{-6}$ precision in $E[V]$ for up to $N=10$. 

Since the evolution times are permutable, we need to avoid covering the identical algorithms. In order to do this the function is minimized over the variables $dt_1$, $dt_2$, ...,$dt_N$, where $dt_1=t_1$ is the first evolution time and $dt_k  = t_{k}-t_{k-1}$. The PSO 'particles' thus have $N$-dimensional coordinates and the 'goodness' of their position is defined as $E[V]$ for these evolution times. If $dt_k<0$ we manually set the value of $E[V]$ to be some large number (instead of calculating the expected variance), creating a barrier for the particles so that they cover only unique algorithms. 

We use a swarm of $16$ particles, randomly initiated with $dt_1 \in (0, 1.5)$  and $dt_i \in (0,2)$, which is a reasonable range implied from the discrete LONA algorithm. The initial velocities are arbitrarily set to $(-1,1)$ range and their absolute values are limited from above by $2$. 

\begin{figure}[!t]
\begin{picture}(400,167)
\put(20,3){\includegraphics[width=230pt]{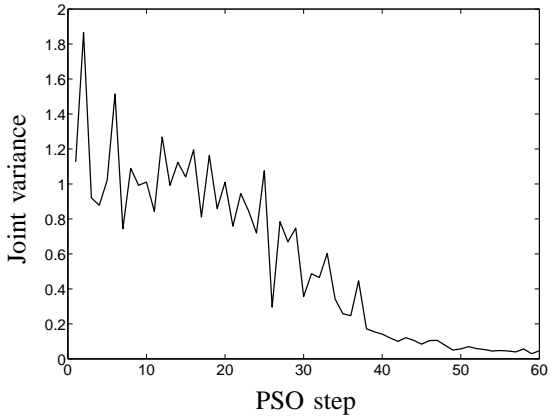}}
\put(27,57){\rotatebox{90}{Joint variance}}
\put(121,2){PSO step}
\end{picture}
\caption{Joint variance convergence with the number of PSO updates for $5$-measurement estimation algorithm.}
\label{jointvariance}
\end{figure}

We employ the velocity updating rule with a constriction factor $\chi$ as follows
\begin{equation}
v_{i+1} = \chi \left(v_i + r_1 c_1  (x_l - x) + r_2 c_2 (x_g - x) \right),
\end{equation}
where $x_l$ and $x_g$ are local and global best coordinates so far and $r_{1,2} \in (0,1)$ are random numbers generated independently for each particle every swarm updating step. The parameters used are $c_1 = c_2 = 2.05$ and 
\begin{equation}
\chi = \frac{2}{\left|2 - \varphi - \sqrt{\varphi ^2 - 4 \varphi}\right|} \approx 0.729,\end{equation} where $\varphi = c_1 + c_2$, as was suggested in \cite{Constriction}. Numeric simulations showed that this technique performs much better than the standard approach with $\chi$ and $c_1= c_2 = 2$, having a relatively quick and stable convergence and no 'swarm explosion' occurrences. 

The optimization was done for $5$ and $10$ measurement steps using the swarm of $16$ particles. For $5$ steps we initiated the coordinates as being random $\pm 0.5$ within LONA algorithm and for $10$ steps $\pm 0.1$. A typical plot for the convergence of joint variance of the swarm is shown on Fig. \ref{jointvariance} for $5$-measurement algorithm. The best algorithms found are shown in Tables \ref{ALG5} and \ref{ALG10} and their performance is shown as crosses on Fig. \ref{LONAfig}.  

\section{Conclusion}
We demonstrated how the performance of qubit Hamiltonian estimation algorithm can be improved using PSO. This approach requires several orders less operations in order to achieve similar precision in evolution times defining the algorithm. Thus, we demonstrated how PSO can be applied to complex quantum algorithms optimization to achieve noticeable results and that it can be useful in such applications. The further advancements would include creating a realistic estimation process model for DQD system in order to optimize the procedure for a real experimental set-up.


\begin{thebibliography}{99}

\bibitem{NC} M. A. Nielsen and I. L. Chuang, \textit{Quantum Computation and Quantum Information}, Cambridge University Press, Cambridge, England, 2000.

\bibitem{Tom1} I. L. Chuang and M. A. Nielsen, \textit{J. Mod. Opt.} {\bf 44}, 2455 (1997).

\bibitem{Tom2} J. F. Poyatos, J. I. Cirac, and P. Zoller, \textit{Phys. Rev. Lett.} {\bf 78}, 390 (1997).

\bibitem{cole} J. H. Cole, S. G. Schirmer, A. D. Greentree, C. J. Wellard, D. K. L. Oi, and L. C. L. Hollenberg,
\textit{Phys. Rev. A} {\bf 71}, 062312 (2005).

\bibitem{QPEA} R. Cleve, A. Ekert, C. Macchiavello, and M. Mosca, \textit{Proc. R. Soc. London A} \textbf{454}, 339 (1998).

\bibitem{Reilly} D. J. Reilly \textit{et al.}, \textit{Science} \textbf{321}, 817 (2008).

\bibitem{Foletti} S. Foletti, H. Bluhm, D. Mahalu, V. Umansky, and A. Yacoby,
\textit{Nat. Phys.} \textbf{5}, 903 (2009).

\bibitem{WeismanNature} B. L. Higgins, D. W. Berry, S. D. Bartlett, H. M. Wiseman, and G. J. Pryde, \textit{Nature} {\bf 450}, 393 (2007).

\bibitem{Sergeevich}
A.~Sergeevich, A.~Chandran, J.~Combes, S.~D.~Bartlett, and H.~M.~Wiseman, \textit{Phys. Rev. A} \textbf{84}, 052315 (2011)

\bibitem{KEPSO} J. Kennedy and R. Eberhart, in \textit{Proc. of the IEEE International Conf. on Neural Networks}, Perth, Australia (1942).

\bibitem{PSObook} J. Kennedy and R.C. Eberhart and Y. Shi, \textit{Swarm intelligence}, Morgan Kaufmann Publishers, San Francisco, 2001.

\bibitem{Hentschel1}
A. Hentschel and B. C. Sanders, \textit{Phys. Rev. Lett.} \textbf{104}, 063603 (2010)

\bibitem{Hentschel2}
A. Hentschel and B. C. Sanders, \textit{Phys. Rev. Lett.} \textbf{107}, 233601 (2011)

\bibitem{WisMilBook} H. M. Wiseman and G. J. Milburn, \textit{Quantum Measurement and Control}, (Cambridge Univ. Press, Cambridge, 2010)  

\bibitem{Constriction} M. Clerc and J. Kennedy, \textit{IEEE Trans. Evolutionary Computation} \textbf{6(1)}, 58 (2002). 

\end{thebibliography}
\end{document}